\documentclass[aps,12pt,pra,showpacs,superscriptaddress,preprint]{revtex4}

\usepackage{amsmath}

\usepackage{graphicx}

\usepackage{subfig}

\usepackage{array}

\usepackage{amsfonts}

\usepackage{epsf}

\usepackage{epsfig}

\usepackage{amssymb}

\usepackage{bbm}

\usepackage{delarray}

\usepackage{caption}

\usepackage{amstext}

\usepackage{graphics}

\usepackage{epstopdf}

\usepackage{color}

\catcode`ð=\active

\defð{\u{g}}

\catcode`Ð=\active

\defÐ{\u{G}}

\catcode`Ý=\active

\defÝ{\. I}

\catcode`ö=\active

\defö{\"{o}}

\catcode`Ö=\active

\defÖ{\"O}

\catcode`ü=\active

\defü{\"{u}}

\catcode`Ü=\active

\defÜ{\"{U}}

\catcode`Þ=\active

\defÞ{\c{S}}

\catcode`þ=\active

\defþ{\c{s}}

\catcode`ý=\active

\defý{{\i}}

\catcode`ç=\active

\defç{\d{c}}

\catcode`Ç=\active

\defÇ{\d{C}}

\begin{document}

\title{Schr{\"{o}}dinger Equation with a Non-Central Potential: Some Statistical Quantities}
\author{\small Altuð Arda}
\altaffiliation[Present adress: ]{Department of Mathematical Science, City University London,
Northampton Square,\\ London EC1V 0HB, UK}\affiliation{Department of
Physics Education, Hacettepe University, 06800, Ankara,Turkey}
\author{\small Cevdet Tezcan}
\affiliation{Faculty of Engineering, Baskent University, Baglica Campus, Ankara,Turkey}
\author{\small Ramazan Sever}
\email[E-mails: ]{arda@hacettepe.edu.tr, ctezcan@baskent.edu.tr, sever@metu.edu.tr}\affiliation{Department of
Physics, Middle East Technical  University, 06531, Ankara,Turkey}

\begin{abstract}
In this paper, we search the dependence of some statistical quantities such as the free energy, the mean energy, the entropy, and the specific heat for the Schrödinger equation on the temperature, particularly the case of a non-central potential. The basic point is to find the partition function which is obtained by a method based on the Euler-Maclaurin formula. At first, we present the analytical results by supporting with some plots for the thermal functions for one- and three-dimensional cases to find out the effect of the angular momentum. We also search then the effect of the angle-dependent part of the non-central potential. We discuss the results briefly for a phase transition for the system. We also present our results for three-dimesional harmonic oscillator.
\end{abstract}


\maketitle

\newpage

\section{Introduction}
The non-central potentials especially for which the Schrödinger equation can be solved exactly by separation of variables have been found many applications, particularly in quantum chemistry [1]. These potentials are important because they represent the nature of non-central forces. They are used to describe the quantum dynamics of ring-shaped molecules, and interactions between deformed nuclei pairs [2]. The potentials without spherical symmetry have some applications within the nanostructure theory [3], and also help us about structuring the metallic glasses [4]. The non-central potentials serve to the theory of the material sciences, for example, describing microscopic elasticity, and obtaining of elastic constants of a cubic crystal [5]. As it is well known that the Makarov [6], and the Hartmann potentials [7], as non-central potentials, have been especially studied by many groups. The exact, analytical solutions of the non-central potentials and their generalizations have been studied in relativistic/non-relativistic regions for many years. The the results are obtained either for the constant mass or for the position-dependent mass formalism [8-14]. In the present work, we restrict ourselves to study the statistical functions for a particle moving in a non-central potential given by [15]
\begin{eqnarray}
V(r,\theta)=a^{2}_{1}r^{2}+\frac{1}{r^{2}}\left(\frac{a^{2}_{2}}{\sin^{2}\theta}+a^{2}_{3}\cot^{2}\theta\right)\,,
\end{eqnarray}
This form corresponds to the potential studied relativistically in Ref. [15] with A = 0, and p = 2.

The thermodynamics functions such as the free energy, the mean energy, the entropy, and the specific heat have been studied in relativistic domain and in non-relativistic one [16-21]. Pacheco and co-workers have firstly studied the one-dimesional Dirac oscillator in a thermal bath to analyze some thermal functions, they have then extended this subject to three-dimensional case [16, 17]. Boumali's group has presented the results about some statistical functions both for the Klein-Gordon and Dirac oscillators, and also for Dirac oscillator in an external magnetic field [18, 20]. The thermodynamical properties of graphene have been studied in Ref. [19], and the thermodynamic quantities have been searched for the Dirac oscillator with Aharonov-Bohm potential in a static electromagnetic field in Ref. [21]. It seems that studying of the statistical quantities receives still a growing interest in literature. In Ref. [22], the relativistic, spin-$1$ Duffin-Kemmer-Petiau oscillator subjected to an external magnetic field in non-commutative space has been studied for thermal functions. In Ref. [23], the statistical functions have been searched for the one-dimensional, spin-$1$ Kemmer equation with Dirac-oscillator potential. The vibrational partition function and other related quantities have been also studied for squared plus inverse-squared potential in non-relativistic domain [24]. Here we prefer, for the first time, to study some thermal functions in the non-relativistic region, particularly for the non-central potential given in Eq. (1). We compute the free energy, the mean energy, the entropy, and the specific heat with the help of the partition function which will be written in terms of a dimensionless parameter. We use a method based on the Euler-MacLaurin formula [16] to complete the numerical analyze.

The paper is organized as follows. In Section II, we first compute the analytical solutions of the Schrödinger equation for the non-central potential where we use the parametric generalization of the Nikiforov-Uvarov method [25]. We outline the basic points of this method in Appendix. In the next subsection, after determining the partition function by using the Euler-MacLaurin formula, we give some statistical functions such as the free energy, the mean energy, the entropy, and the specific heat. In last Section, we give the conclusion.

\section{The Bound States \lowercase{and} Statistical Quantities}

\subsection{B\lowercase{ound} S\lowercase{tates}}

The time-independent Schrödinger equation is written in spherical coordinates as [26]
\begin{eqnarray}
\left\{-\frac{\hbar^{2}}{2M}\vec{\nabla}^{2}+V(r, \theta, \phi)-E\right\}\Psi(r, \theta, \phi)=0\,,
\end{eqnarray}
with
\begin{eqnarray}
\vec{\nabla}^{2}=\frac{1}{r^2}\frac{\partial}{\partial r}\left(r^{2}\frac{\partial}{\partial r}\right)+\frac{1}{r^2}\left[\frac{1}{\text{sin}\theta}\frac{\partial}{\partial\theta}\left(\text{sin}\theta\frac{\partial}{\partial\theta}\right)+\frac{1}{\text{sin}^{2}\theta}\frac{\partial^{2}}{\partial\phi^{2}}\right]\,.\nonumber
\end{eqnarray}
In Eq. (2), $M$ is the mass and $E$ is the energy of the particle.

Writing the total wave function with magnetic quantum number $m$ as
\begin{eqnarray}
\Psi(r, \theta, \phi)=R(r)\Theta(\theta)e^{\mp im\phi}\,\,; R(r)=\frac{f(r)}{r}\,\,; m=0, 1, 2, \ldots
\end{eqnarray}
and inserting the non-central potential into Eq. (2) gives two differential equations for coordinates of $\theta$ and $r$
\begin{eqnarray}
\frac{d^2\Theta(\theta)}{d\theta^{2}}+\frac{\text{cos}\theta}{\text{sin}\theta}\frac{d\Theta(\theta)}{d\theta}+\left[\ell(\ell+1)-\frac{m^{2}+\frac{2M}{\hbar^2}a^{2}_{2}}
{\text{sin}^{2}\theta}-\frac{2Ma^{2}_{3}}{\hbar^2}\cot^{2}\theta\right]\Theta(\theta)=0\,.
\end{eqnarray}
and
\begin{eqnarray}
\frac{d^{2}f(r)}{dr^{2}}+\frac{2M}{\hbar^{2}}\left[E-a^{2}_{1}r^{2}-\frac{\ell(\ell+1)\hbar^{2}}{2Mr^{2}}\right]f(r)=0\,,
\end{eqnarray}
where $\ell$ is the angular momentum quantum number.

Defining a new variable as $y=1+\text{cos}\,\theta$ converts Eq. (4) into following one
\begin{eqnarray}
\frac{d^2\Theta(y)}{dy^2}&-&\frac{1}{\left[y(1-\frac{y}{2})\right]^2}\left[\frac{1}{4}(m^2+\frac{2Ma^{2}_{2}}{\hbar^2})+\frac{2Ma^{2}_{3}}{\hbar^2}
-\left(\frac{1}{2}\ell(\ell+1)+\frac{2Ma^{2}_{3}}{2\hbar^2}\right)y\right.\nonumber\\&&\left.+\left(\frac{1}{4}\ell(\ell+1)+\frac{2Ma^{2}_{3}}{4\hbar^2}\right)y^2\right]\Theta(y)=0\,,
\end{eqnarray}
At this point, we start to follow the steps of the Nikiforov-Uvarov method summarized in Appendix to find the analytical solutions. Eq. (A.2) in Appendix gives a relation connecting the angular momentum quantum number to other potential parameters giving as $\ell=\left[L+1/2\right]$ with $s=0, 1, 2, \ldots$. The corresponding solution functions from (A.4) in Appendix
\begin{eqnarray}
\Theta(y) \sim y^{1+\Lambda}(1-y)^{\Lambda}P_{s}^{(\Lambda, \Lambda)}(1-y)\,.
\end{eqnarray}
 They are expressed in terms of the Jacobi polynomials, $P_{s}^{(\Lambda, \Lambda)}(y)$, with $\Lambda=\sqrt{1+m^2+\frac{2M}{\hbar^2}(a^{2}_{2}+a^{2}_{3})\,}$, and $L=-1+\frac{1}{2}\sqrt{(1+2s+2\Lambda)^2-\frac{8Ma^{2}_{3}}{\hbar^2}\,}$ in the above.

To solve Eq. (5) we use a new variable as $y=\frac{1}{\hbar}\sqrt{2m a^{2}_{1}\,}r^{2}$
\begin{eqnarray}
y\frac{d^2f(y)}{dy^2}+\frac{1}{2}\frac{df(y)}{dy}+\frac{1}{4}\left(\frac{\sqrt{2M\,}E}{\hbar a_{1}}-y-\frac{\ell(\ell+1)}{y}\right)f(y)=0\,
\end{eqnarray}
In order to obtain the energy spectrum to the above equation, we have to analyze the asymptotic behavior of the wave function for $r \rightarrow 0$ and $r \rightarrow \infty$. So, it is convenient to write it in the following form
\begin{eqnarray}
f(y)=y^{\mu}e^{-y/2}h(y)\,,
\end{eqnarray}
with $\mu=(1/2)(\ell+1)$. By substituting Eq. (9) into Eq. (8) we obtain
\begin{eqnarray}
\frac{d^2h(y)}{dy^2}+\frac{2\mu+\frac{1}{2}-y}{y}\frac{dh(y)}{dy}-\left[\frac{\mu+\frac{1}{4}-\frac{\sqrt{2M\,}E}{4\hbar a_{1}}}{y^{2}}\right]h(y)=0\,,
\end{eqnarray}
and its solutions can be given from (A.7) in Appendix as
\begin{eqnarray}
h(y) \sim \frac{\Gamma(n+\frac{3}{2}+\ell)}{n!\Gamma(\frac{3}{2}+\ell)} \,_{1}F_{1}(-n; \frac{3}{2}+\ell; y)\,.
\end{eqnarray}
The required quantization condition (A.2) gives us
\begin{eqnarray}
\mu+\frac{1}{4}-\frac{\sqrt{2M\,}E}{4\hbar a_{1}}=-n\,;\,\,\,n=0, 1, 2, \ldots\,.
\end{eqnarray}

With the help of last equation, we write the bound state solutions of the present problem more clearly
\begin{eqnarray}
E_{n\ell}=\frac{\hbar a_{1}}{\sqrt{2M\,}}(4n+2\ell+3)\,.
\end{eqnarray}

From Eq. (3) and with the help of Eqs. (7) and (11), the total wave functions can be written as
\begin{eqnarray}
\Psi(r, \theta, \phi) \sim y^{(1/2)(\ell+2\Lambda+3)}(1-y)^{\Lambda}e^{-y/2}\,_{1}F_{1}(-n; 2\mu+\frac{1}{2}; y)P_{s}^{(\Lambda, \Lambda)}(1-y)e^{\mp im\phi}\,,
\end{eqnarray}
where it should be stressed that the special case of Jacobi polynomials $P_{n}^{(\sigma_{1}, \sigma_{2})}(y)$ such as $\sigma_{1}=\sigma_{2}$ is given in terms of the Gegenbauer polynomials [27].

In the next section, we will firstly perform the partition function of the whole system by using energy eigenvalues given in Eq. (13).

\subsection{S\lowercase{ome} S\lowercase{tatistical} Q\lowercase{uantities}}
Our starting point is the partition function of the system which is defined as a summation all over the quantum states [16, 17]
\begin{eqnarray}
Z(\beta)=\sum_{n'=0}^{\infty}\omega(E_{n'})e^{-\beta(E_{n'}-E_{0})}=e^{\beta E_{0}}\sum_{n'=0}^{\infty}\omega(E_{n'})e^{-\beta E_{n'}}\,,
\end{eqnarray}
where $\beta=1/k_{B}T$, $k_{B}$ Boltzmann constant, and $T$ is temperature in Kelvin. The factor of $\omega(E_{n'})$ is the degree of degeneracy for the quantum level given as
\begin{eqnarray}
E_{n'}=\sqrt{\frac{\hbar^{2}a^{2}_{1}}{2M}\,}(2n'+3)\,,
\end{eqnarray}
with $n'=N+\ell$, $n' \geq 0$ where we introduce a new 'quantum number' as $N=2n$. For each quantum level with $(n, \ell)$ there are $2\ell+1$ degenerate states differing with values of magnetic quantum number $m$. For a given $n'$, the total degree of degeneracy is obtained as
\begin{eqnarray}
\sum_{\ell=0}^{n'}(2\ell+1)=(1+n')^{2}\,.
\end{eqnarray}

As a result, the partition function becomes
\begin{eqnarray}
Z(\beta)=e^{3\beta\xi}\sum_{n'=0}^{\infty}(1+n')^{2}e^{-\beta\xi(2n'+3)}\,,
\end{eqnarray}
with $\xi=\sqrt{\frac{\hbar^{2}a^{2}_{1}}{2M}\,}$.

We study the following thermal quantities such as the Helmholtz free energy, the mean energy, the entropy, and the specific heat defined in terms of the partition function as following [16, 17]
\begin{eqnarray}
F(\beta)=-\frac{1}{\beta}\,\text{ln}\,Z(\beta); U(\beta)=-\frac{\partial}{\partial\beta}\,\text{ln}\,Z(\beta); S(\beta)=k_{B}\beta^2\frac{\partial}{\partial\beta}F(\beta); C(\beta)=-k_{B}\beta^2\frac{\partial}{\partial\beta}U(\beta)\,,\nonumber\\
\end{eqnarray}
which can be written in terms of a new dimensionless parameter $\bar{\alpha}=1/\beta\xi$
\begin{eqnarray}
&&\bar{F}=F/\xi=-\bar{\alpha}\text{ln}\,Z(\bar{\alpha})\,,\nonumber\\
&&\bar{U}=U/\xi=\bar{\alpha}^{2}\frac{\partial}{\partial\bar{\alpha}}\,\text{ln}\,Z(\bar{\alpha})\,,\nonumber\\
&&\bar{S}=S/k_{B}=\text{ln}\,Z(\bar{\alpha})+\bar{\alpha}\frac{\partial}{\partial\bar{\alpha}}\,\text{ln}\,Z(\bar{\alpha})\,,\nonumber\\
&&\bar{C}=C/k_{B}=2\bar{\alpha}\frac{\partial}{\partial\bar{\alpha}}\,\text{ln}\,Z(\bar{\alpha})+\bar{\alpha}^{2}\frac{\partial^{2}}{\partial\bar{\alpha}^{2}}\,\text{ln}\,Z(\bar{\alpha})\,.
\end{eqnarray}

The following associated integral
\begin{eqnarray}
\int_{0}^{\infty}(1+x)^{2}e^{-\beta\xi(2x+3)}dx=\frac{1}{4\beta^{3}\xi^{3}}[1+2\beta\xi(1+\beta\xi)]e^{-3\beta\xi}\,,
\end{eqnarray}
shows that the series in Eq. (18) is convergent [16]. Eq. (21) makes it possible to employ the Euler-MacLaurin formula [16, 28] for evaluating the partition function numerically
\begin{eqnarray}
\sum_{m=0}^{\infty}f(m)=\frac{1}{2}f(0)+\int_{0}^{\infty}f(x)dx-\sum_{k=1}^{\infty}\frac{1}{(2k)!}B_{2k}f^{(2k-1)}(0)\,,
\end{eqnarray}
where $B_{2k}$ are the Bernoulli numbers given as $B_{2}=1/6$, $B_{4}=-1/30$, $\ldots$ [16]. Up to $k=2$, Eqs. (18) and (21) give the partition function of the system as
\begin{eqnarray}
Z(\bar{\alpha})=\frac{1}{3}+\frac{\bar{\alpha}^{3}}{4}\left[1+\frac{2}{\bar{\alpha}}\left(1+\frac{1}{\bar{\alpha}}\right)\right]+\frac{1}{20\bar{\alpha}}\left[3+\frac{2}{3\bar{\alpha}}\left(1-\frac{1}{3\bar{\alpha}}\right)\right]\,.
\end{eqnarray}

Let us first give the results for high temperatures, $\beta \ll 1$. Only the first term in first parenthesis of Eq. (23) gives a significant contribution to partition function. Hence, we have for high-temperature regime
\begin{eqnarray}
Z(\bar{\alpha}) \sim \frac{\bar{\alpha}^{3}}{4}\,;\,\,\,\bar{U}(\bar{\alpha}) \sim 3\bar{\alpha}\,;\,\,\,\bar{C}(\bar{\alpha}) \sim 3.
\end{eqnarray}

The upper limit for the specific heat is seen in Fig. (4) where we plot its variation versus the temperature $T$. The specific heat increases linearly up to this high-temperature value. The behaviour of the free energy versus temperature is nearly linear but decreases while the temperature increases (Fig. (1)). From Fig. (2), we observe that the mean energy has an increasingly behaviour with increasing temperature. Fig. (3) shows that the entropy increases while the temperature increases. The graphs of specific heat (Figs. (4) and (5d)) show that it does not change discontinuously at any value of temperature. So, we may conclude that the system under consideration has no mark about the phase transition. Because, according to the Ehrenfest's classification, for a physical system having a first-order phase transition, the specific heat changes discontinuously at the critical temperature value, and has an infinite peak at this point while in a second-order transition the specific heat changes discontinuously having a finite value [28]. Finally, we see that the method used here based on the use of the Euler-MacLaurin formula is a suitable tool to perform the thermal quantities for the Schrödinger equation with a non-central potential.

At this point, we give the partition function for the ground state to see the effect of the angular momentum on the above statistical quantities. In order to get numerical results, one has to reduce the partition function in Eq. (18) as
\begin{eqnarray}
Z(\beta)=e^{3\beta\xi/2}\sum_{N=0}^{\infty}e^{-\beta\xi(N+\frac{3}{2}\,)}\,,
\end{eqnarray}
where the degree of degeneracy was taken one. Following the same steps, we write the partition function for one-dimensional case
\begin{eqnarray}
Z(\beta)=\frac{1}{2}+\bar{\alpha}+\frac{1}{12\bar{\alpha}}-\frac{\bar{\alpha}^3}{5400}\,.
\end{eqnarray}
The results giving the above expression are shown in Figs. (5a)-(5d). Comparing these with Figs. (1)-(4) show that the variations of the statistical quantities behave in the same general way of their three-dimensional counterpart. In all cases the values of the functions are reduced with respect to the three-dimensional case due to the effect of angular momentum. The variations of the free energy and mean energy are similar to each other. The limiting value of specific heat is three times smaller than that of the three-dimensional case for high temperatures. The effect of the degeneracy on the general behaviour of the system is seen more clearly.

Now let us discuss the effect of the angle-dependent part of the potential briefly for which  we give some analytical expressions. As seen in Eq. (1), the angle-dependent part of potential has two terms whose contributions are given by the constants $a_2$, and $a_3$. For the dependency on the parameter $a_2$, we write the energy eigenvalue equation
\begin{eqnarray}
E=\sqrt{\frac{a^{2}_{1}\hbar^2}{2M}\,}[2(N+\ell)+3]\,,
\end{eqnarray}
where we have to modify $\Lambda$ and $L$ in Eq. (7) as $\Lambda=\sqrt{1+m^2+\frac{2Ma^{2}_{2}}{\hbar^2}\,}$, and $L=-\frac{1}{2}+\Lambda+s$, respectively. On the other hand, for $a_{3}=0$, we have to write $\Lambda=\sqrt{1+m^2+\frac{2Ma^{2}_{3}}{\hbar^2}\,}$, and $L=-1+\frac{1}{2}\sqrt{(1+2\Lambda+2s)^2-\frac{8Ma^{2}_{3}}{\hbar^2}\,}$ in Eq. (27). It seems that the contribution of $a_{3}$ is greater than that of the parameter $a_{2}$. In order to complete the results, we give our analytical results for the radial part of the potential in Eq. (1) corresponding to the harmonic oscillator obtained by $a_2=a_3=0$. The energy expression is
\begin{eqnarray}
E=\sqrt{\frac{a^{2}_{1}\hbar^2}{2M}\,}(4n+2\ell+3)\,,
\end{eqnarray}
where it should be as $\Lambda=\sqrt{1+m^2\,}$, and $\ell=[\Lambda+s]$.

\section{Conclusions}

We have analyzed the thermal functions of the Schrödinger equation, particularly, for the non-central potential stated in Eq. (1). All basic statistical quantities have been evaluated by a method based on the Euler-MacLaurin formula, and the results for high temperatures are also given for the mean energy, and the specific heat. The mean energy increases for increasing $T$ while the free energy decreases with increasing temperature. The entropy has an increasingly behavior for increasing temperature. The specific heat has an increasing behaviour up to an upper value as given in Eq. (24). We have also studied the statistical functions for the one-dimensional case with $\ell=0$. To compare all results between these cases we have plotted the variations all of thermal functions for the one-dimensional case. We have seen the effect of degeneracy on them clearly which has been appeared in specific heat especially. We have discussed the effect of the non-central part of the potential, and the results for the harmonic oscillator part are given.

\section{Acknowledgments}
One of authors (A.A.) thanks Prof Dr Andreas Fring from City University London and the Department of Mathematics for hospitality. This research was partially supported by the Scientific and Technical Research Council of Turkey and through a fund provided by University of Hacettepe.

The authors would like to thank the editor and the reviewer for their kind and valuable suggestions.

\appendix

\section{}

The general form of a second order differential equation which is solved by using the parametric generalization of the Nikiforov-Uvarov method [25] is
\begin{eqnarray}
\frac{d^{2}F(y)}{dy^{2}}+\frac{\beta_{1}-\beta_{2}y}{y(1-\beta_{3}y)}\frac{dF(y)}{dy}-\frac{\xi_{1}y^{2}-\xi_{2}y+\xi_{3}}{[y(1-\beta_{3}y)]^{2}}\,F(y)=0\,,
\end{eqnarray}
with the quantization rule
\begin{eqnarray}
\beta_{2}s-(2s+1)\beta_{5}+(2s+1)(\sqrt{\beta_{9}\,}+\beta_{3}\sqrt{\beta_{8}\,})+s(s-1)\beta_{3}+\beta_{7}+2\beta_{3}\beta_{8}+2\sqrt{\beta_{8}\beta_{9}\,}=0\,,\nonumber\\
\end{eqnarray}
where $s=0, 1, 2, \ldots$.

The parameters $\beta_{i}'s$ within this approach are defined as
\begin{eqnarray}
&&\beta_{4}=\frac{1}{2}(1-\beta_{1});\,\,\beta_{5}=\frac{1}{2}(\beta_{2}-2\beta_{3});\,\,\beta_{6}=\beta^{2}_{5}+\xi_{1};\,\,\beta_{7}=2\beta_{4}\beta_{5}-\xi_{2};\nonumber\\&&\beta_{8}=\beta^{2}_{4}+\xi_{3};\,\,\beta_{9}=\beta_{3}(\beta_{7}+\beta_{3}\beta_{8})+\beta_{6}\,,
\end{eqnarray}

The corresponding wave functions are given in terms of the parameters  $\beta_{i}$ [25]
\begin{eqnarray}
F(y) \sim y^{\beta_{12}}(1-\beta_{3}y)^{-\beta_{12}-\frac{\beta_{13}}{\beta_{3}}}\,P_{s}^{(\beta_{10}-1,\,\frac{\beta_{11}}{\beta_{3}}-\beta_{10}-1)}(1-2\beta_{3}y)\,.
\end{eqnarray}
where
\begin{eqnarray}
&&\beta_{10}=\beta_{1}+2\beta_{4}+2\sqrt{\beta_{8}\,};\,\,\beta_{11}=\beta_{2}-2\beta_{5}+2(\sqrt{\beta_{9}\,}+\beta_{3}\sqrt{\beta_{8}\,});\nonumber\\&&\beta_{12}=\beta_{4}+\sqrt{\beta_{8}\,};\,\,\beta_{13}=\beta_{5}-(\sqrt{\beta_{9}\,}+\beta_{3}\sqrt{\beta_{8}\,})\,.
\end{eqnarray}
with the Jacobi polynomials $P_{n}^{(\sigma_{1}, \sigma_{2})}(y)$.

If the parameter $\beta_{3}$ is zero, then the quantization condition changes into
\begin{eqnarray}
\beta_{2}s+(1-2s)\beta_{5}+(2s+1)(\sqrt{\beta_{9}\,}-\beta_{3}\sqrt{\beta_{8}\,})+s(s-1)\beta_{3}+\beta_{7}+2\beta_{3}\beta_{8}-2\sqrt{\beta_{8}\beta_{9}\,}=0\,,\nonumber\\
\end{eqnarray}
with the corresponding wave functions
\begin{eqnarray}
F(y) \sim y^{\beta^{*}_{12}}(1-\beta_{3}y)^{-\beta^{*}_{12}-\frac{\beta^{*}_{13}}{\beta_{3}}}\,P_{s}^{(\beta^{*}_{10}-1,\,\frac{\beta^{*}_{11}}{\beta_{3}}-\beta_{10}-1)}(1-2\beta_{3}y)\,,
\end{eqnarray}
where
\begin{eqnarray}
&&\beta^{*}_{10}=\beta_{1}+2\beta_{4}-2\sqrt{\beta_{8}\,};\,\,\beta^{*}_{11}=\beta_{2}-2\beta_{5}-2(\sqrt{\beta_{9}\,}-\beta_{3}\sqrt{\beta_{8}\,})\,,\nonumber\\&&\beta^{*}_{12}=\beta_{4}-\sqrt{\beta_{8}\,};\,\,\beta^{*}_{13}=\beta_{5}-(\sqrt{\beta_{9}\,}-\beta_{3}\sqrt{\beta_{8}\,})\,.
\end{eqnarray}

\newpage

\begin{figure}
\centering
\includegraphics[height=3.5in, width=5in, angle=0]{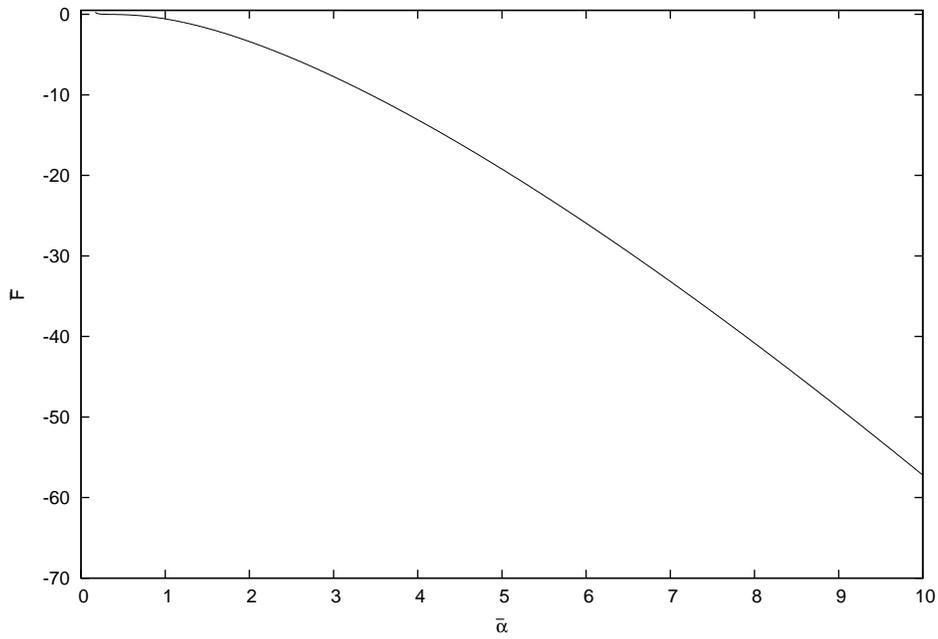}
\caption{The variation of the free energy versus $\bar{\alpha}$.}
\end{figure}

\begin{figure}
\centering
\includegraphics[height=3.5in, width=5in, angle=0]{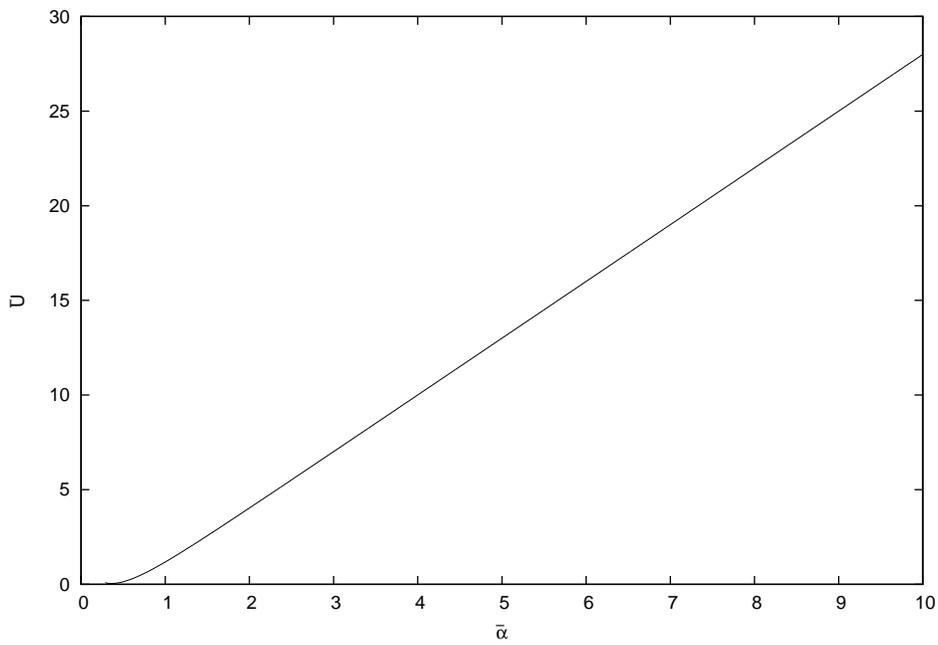}
\caption{The variation of the mean energy versus $\bar{\alpha}$.}
\end{figure}

\begin{figure}
\centering
\includegraphics[height=3.5in, width=5in, angle=0]{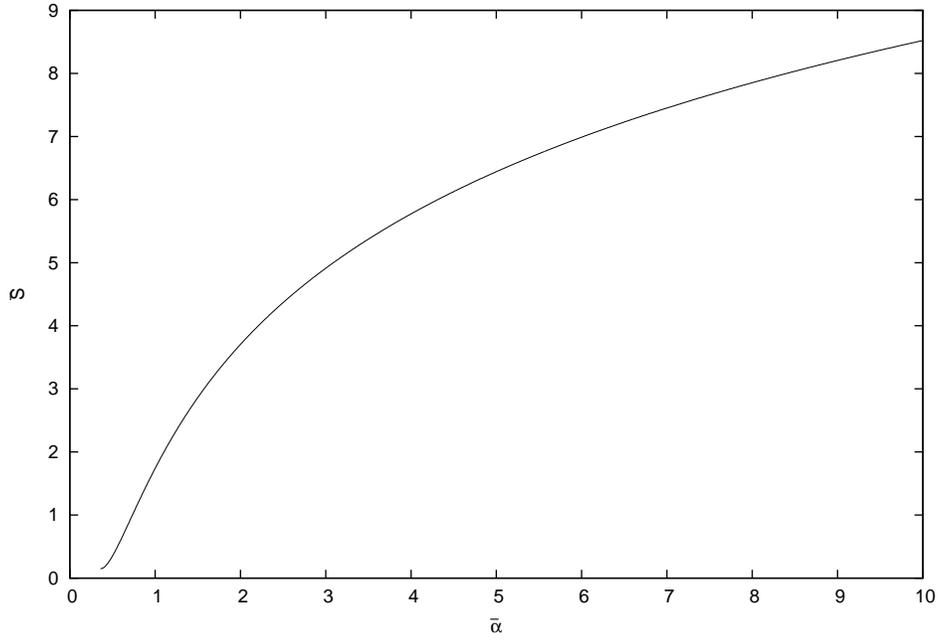}
\caption{The variation of the entropy versus $\bar{\alpha}$.}
\end{figure}

\newpage

\begin{figure}
\centering \subfloat[][The specific heat for nearly low temperatures.]{\includegraphics[height=2.2in,
width=3in, angle=0]{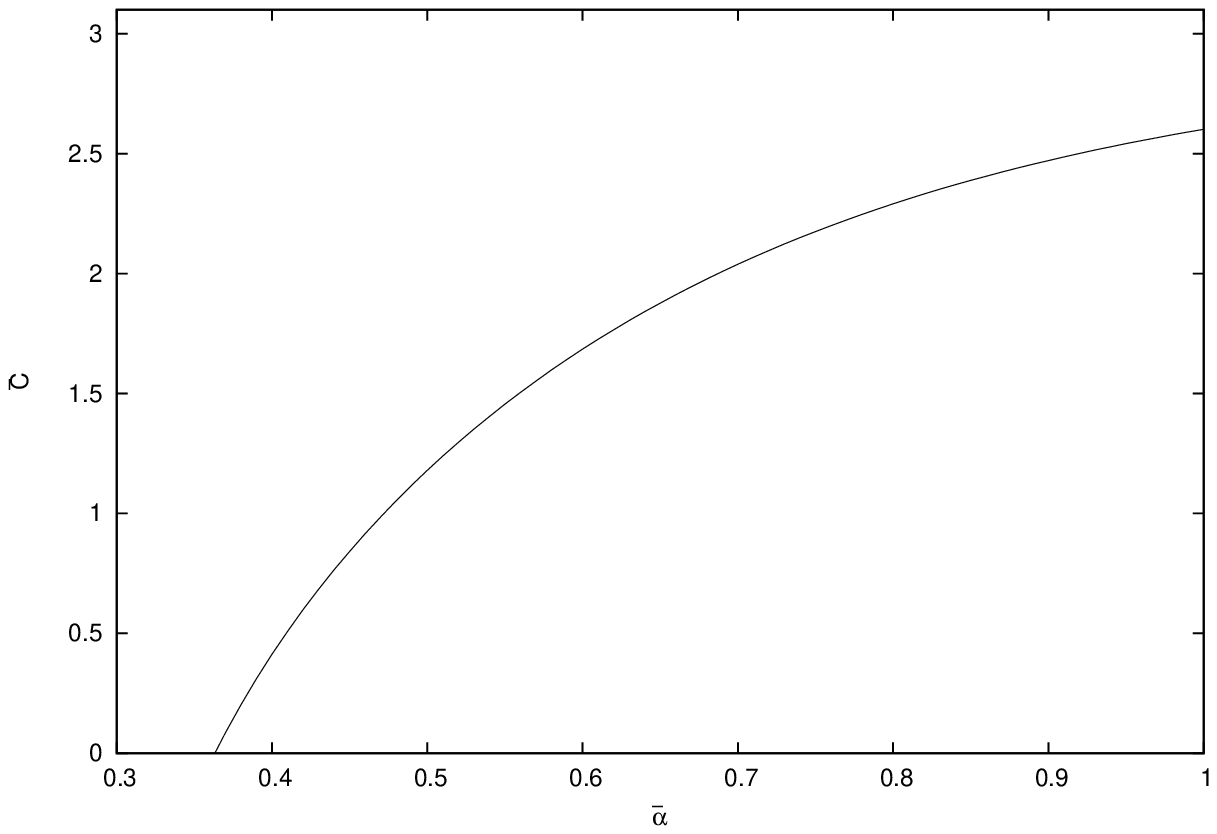}}
\subfloat[][The specific heat for high temperatures.]{\includegraphics[height=2.2in, width=3in,
angle=0]{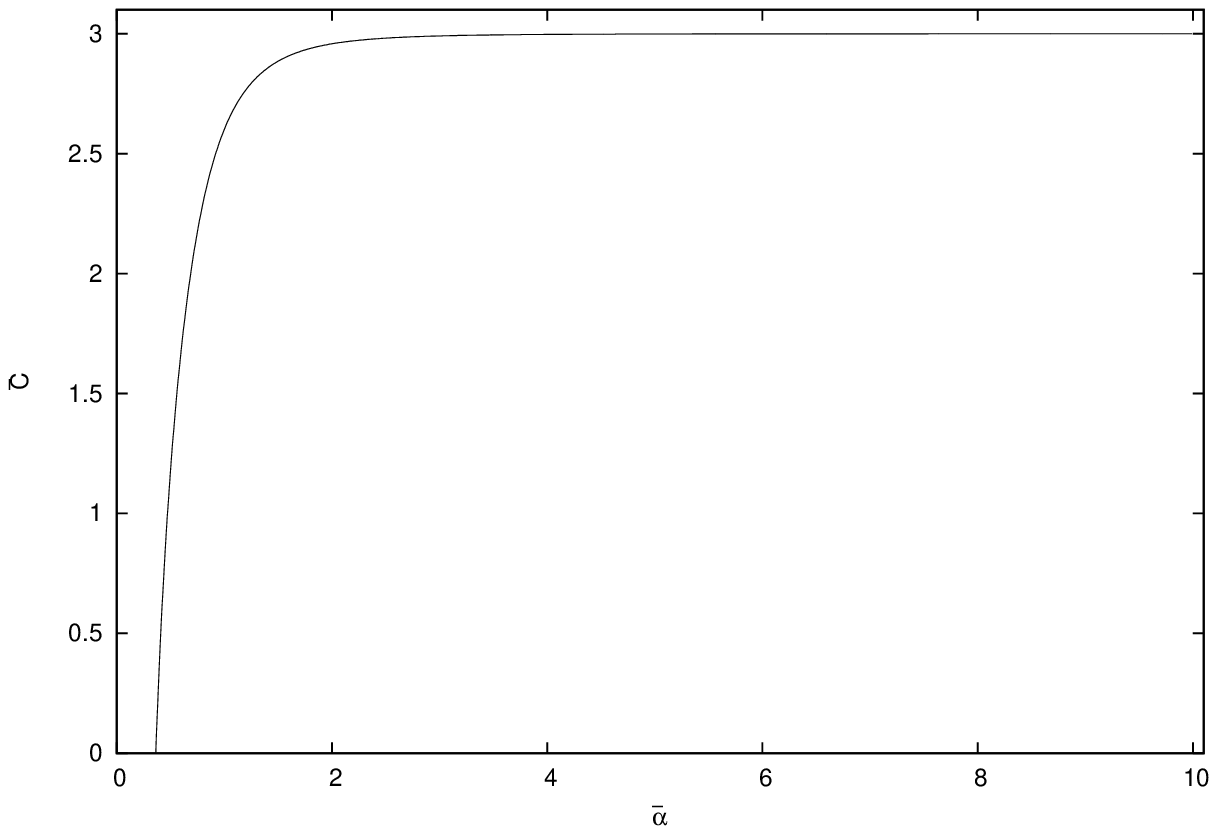}}
\caption{The variation of the specific heat versus $\bar{\alpha}$.}
\end{figure}

\begin{figure}
\centering \subfloat[][The dependence of free energy on $\bar{\alpha}$.]{\includegraphics[height=2.2in,
width=3in, angle=0]{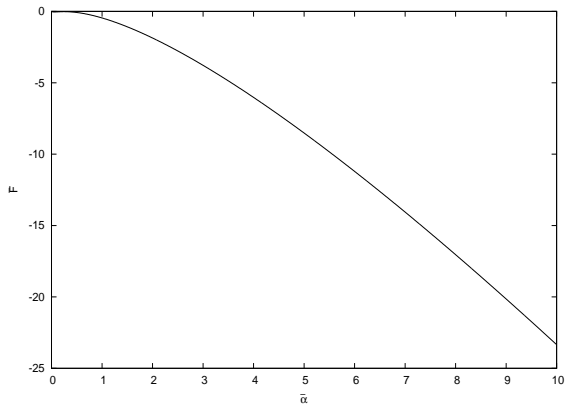}}
\subfloat[][The dependence of mean energy on $\bar{\alpha}$.]{\includegraphics[height=2.2in, width=3in,
angle=0]{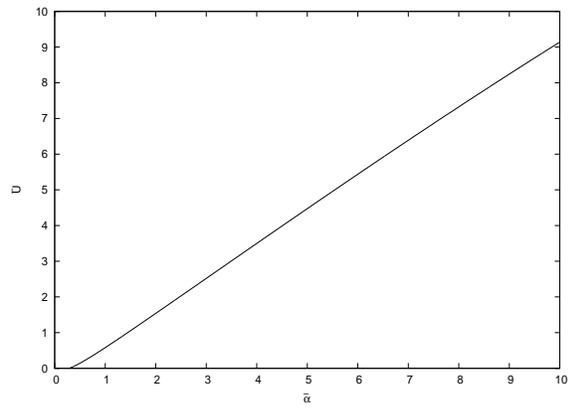}}\\
\subfloat[][The dependence of entropy on $\bar{\alpha}$.]{\includegraphics[height=2.2in, width=3in,
angle=0]{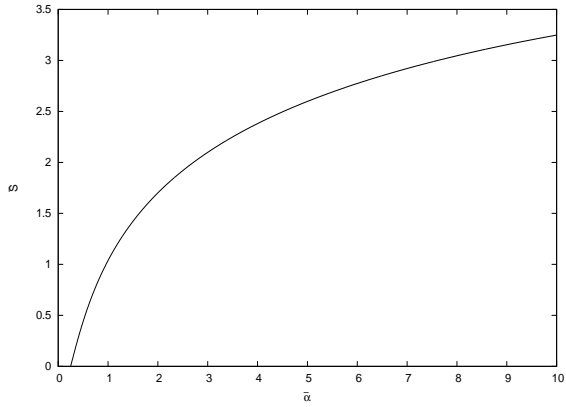}}
\subfloat[][The dependence of specific heat on $\bar{\alpha}$.]{\includegraphics[height=2.2in, width=3in,
angle=0]{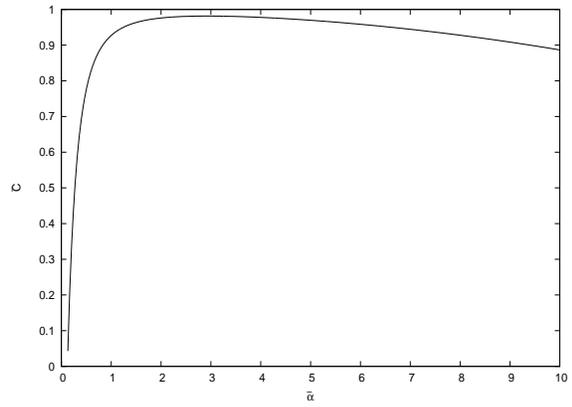}}
\caption{The variation of statistical quantities versus $\bar{\alpha}$ for the one-dimensional case.}
\end{figure}

\end{document}